\documentclass[twocolumn,prb,aps,floatfix,amsmath,amssymb,showpacs]{revtex4}

\usepackage{graphicx}% Include figure files
\usepackage{dcolumn}% Align table columns on decimal point
\usepackage{bm}% bold math

\begin{document}

\preprint{draft - do not distribute}

\title{Excitations and quantum fluctuations in site diluted two-dimensional
antiferromagnets}

\author{Eduardo R. Mucciolo,$^1$ A. H. Castro Neto,$^2$ and
Claudio Chamon$^2$}

\affiliation{$^1$Department of Physics, University of Central Florida,
Orlando, Florida 32816, USA;\\Department of Physics, Duke University,
Durham, North Carolina 27708, USA and\\Departamento de F\a'{\i}sica,
Pontif\a'{\i}cia Universidade Cat\'olica do Rio de Janeiro, Caixa
Postal 37801, 22452-970 Rio de Janeiro, Brazil\\$^1$Department of
Physics, Boston University, Boston, Massachusetts 02215, USA}

\date{\today}

\begin{abstract}
We study the effect of site dilution and quantum fluctuations in an
antiferromagnetic spin system on a square lattice within the linear
spin-wave approximation. By performing numerical diagonalization in
real space and finite-size scaling, we characterize the nature of the
low-energy spin excitations for different dilution fractions up to the
classical percolation threshold. We find nontrivial signatures of
fractonlike excitations at high frequencies. Our simulations also
confirm the existence of an upper bound for the amount of quantum
fluctuations in the ground state of the system, leading to the
persistence of long-range order up to the percolation threshold. This
result is in agreement with recent neutron-scattering experimental
data and quantum Monte Carlo numerical calculations. We also show that
the absence of a quantum critical point below the classical
percolation threshold holds for a large class of systems whose
Hamiltonians can be mapped onto a system of coupled noninteracting
massless bosons.
\end{abstract}


% 75.30.Ds Spin waves
% 75.40.Mg Numerical simulation studies
% 75.50.Ee Antiferromagnetics
% 75.10.Jm Quantized spin models
% 75.10.Nr Spin-glass and other random models
% 75.40.Cx Static properties (order parameter, static susceptibility, 
%          heat capacities, critical exponents, etc.)

%\keywords{disorder, antiferromagnetism, spin waves}

\maketitle

%%%%%%%%%%%%%%%%%%%%%%%%%%%%%%%%%%%%%%%%%%%%%%%%%%%%%%%%%%%%%%%%%%%%%%%%%
%%%%%%%%%%%%%%%%%%%%%%%%%%%%%%%%%%%%%%%%%%%%%%%%%%%%%%%%%%%%%%%%%%%%%%%%%
\section{Introduction}
\label{sec:intro}

The problem of the interplay of quantum fluctuations and disorder in
low dimensional systems is of fundamental importance in modern
condensed matter physics. It is relevant for the understanding of the
metal-insulator transition in metal-oxide-semiconductor field-effect
transistors (MOSFETS),\cite{serguei} impurity effects in $d$-wave
superconductors,\cite{dwave} non-Fermi-liquid behavior in U and Ce
intermetallics,\cite{heavy-fermions} and the persistence of long-range
order (LRO) in two-dimensional (2D) spin-$1/2$ quantum
antiferromagnets (AFM),\cite{vajk02science} among others.

The 2D square lattice with nearest-neighbor hopping undergoes a
classical percolation transition upon random dilution. For the case of
site dilution, the transition occurs at the hole concentration $x_c
\approx 0.41$,\cite{newman00} while for bond dilution the largest
(infinite) connected cluster disappears when exactly half of the bonds
are absent (i.e., $x_c = 1/2$).\cite{nakayama94, stauffer94} Thus, no
ground-state long-range order is possible for any model with short
range interactions in these lattices above $x_c$. For those models
where order does exist in the clean limit, it is natural to ask
whether dilution can enhance quantum fluctuations to the point of
destroying long-range order at some doping fraction below $x_c$. This
possibility has led to several theoretical and experimental
investigations in a variety of systems.\cite{vajk02science,
sandvik02prb, sasha, senthil, Ali-Moore} In particular, a recent
neutron scattering experiment \cite{vajk02science} in the site-diluted
$S=1/2$ Heisenberg quantum AFM La$_2$Cu$_{1-x}$(Zn,Mg)$_x$O$_4$
(LCMO), indicates that LRO exists up to $x_c$. This fact was
corroborated by an extensive quantum Monte Carlo (QMC) simulation
\cite{sandvik02prb} and by spin-wave theory (SWT) analytical
calculation in the dilute regime.\cite{sasha} The numerical and
experimental data suggest that the disappearance of order in the
ground state is dominated by a classical effect and no quantum phase
transition takes place below $x_c$. The analytical calculation,
\cite{sasha} on the other hand, points to a nontrivial
dilution-induced softening of low-frequency spin-wave
excitations. However, the latter was carried out within the $T$-matrix
approximation, which excludes coherent superposition and interference,
and thus could not account for localization effects.

In this work we carry out exact real space numerical diagonalizations
of the dilute 2D Heisenberg AFM within the linear spin-wave
approximation for dilution fractions ranging from the clean limit to
the the classical percolation threshold. Although our method cannot be
used to investigate systems as large as those used in previous
numerical studies of the dynamical structure factor
alon,\cite{terao94} it provides complete access to eigenergies and
eigenvectors, thus allowing us to probe more carefully into the
structure of the excited states of the site dilute AFM. We find that
excitations in this system break up into different modes as the amount
of dilution is increased. The multipeak structure of the spectral
function shows the simultaneous existence of extended (magnons) and
localized (fractons) excitations, similarly to what was observed
experimentally by Uemura and Birgeneau for dilute {\it
three-dimensional} AFMs several years ago.\cite{uemura87}

We also argue that the absence of a quantum critical point in diluted
systems below $x_c$ is a universal feature for a large class of
continuous models of the Heisenberg type that can be mapped into a
system of coupled harmonic oscillators within some approximate
scheme. We establish an upper bound for the amount of quantum
fluctuations in these diluted noninteracting bosonic systems and show
that quantum fluctuations are bounded even at the percolation
threshold for 2D systems. Indeed, one may wonder whether this generic
behavior of diluted bosonic systems is related to universality classes
dictated solely by symmetries of the disordered Hamiltonian, as in the
case of fermionic models.

The paper is organized as follows. In Sec. \ref{sec:hamiltonians} we
define the system Hamiltonian and find the corresponding non-Hermitian
eigenvalue matrix problem in lattice coordinates within the linear
spin-wave approximation. In Sec. \ref{sec:numericalmethod} we present
and discuss the numerical methods used to generate the random
dilution, the identification of the largest connected cluster, and the
diagonalization of the eigenvalue problem. The numerical results for
the spin-wave excitation spectrum and the quantum fluctuation
corrections to the AFM ground state are presented and discussed in
Sec. \ref{sec:numericalresults}. In Sec. \ref{sec:bosons} we argue
that the absence of quantum phase transitions in site dilute systems
occurs for spin Hamiltonians that can be mapped onto a system of
coupled harmonic oscillators. Finally, in Sec. \ref{sec:conclusions},
we draw our conclusions and point to future directions.

%%%%%%%%%%%%%%%%%%%%%%%%%%%%%%%%%%%%%%%%%%%%%%%%%%%%%%%%%%%%%%%%%%%%%%%%%
%%%%%%%%%%%%%%%%%%%%%%%%%%%%%%%%%%%%%%%%%%%%%%%%%%%%%%%%%%%%%%%%%%%%%%%%%
\section{Dilute Heisenberg Antiferromagnet in the Spin Wave Approximation}
\label{sec:hamiltonians}

We begin by reviewing the well-known connection between magnetic and
bosonic systems. In particular, we are interested in spin Hamiltonians
of the form
\begin{eqnarray}
H = \sum_{\langle i, j \rangle,a} \eta_i \eta_j\,  J_a\, S^a_i S^a_j ,
\end{eqnarray}
where $S^a_i$ is the $a=x,y,z$ component of a spin $S$ at site $i$,
$J_a$ is the nearest-neighbor exchange constant, and $\eta_i = 0\,
(1)$ if the site is empty (occupied). The empty sites are randomly
distributed over the whole sample with uniform probability. Calling
$N$ the total number of sites, we define the fraction of occupied
sites as
\begin{equation}
p = \frac{1}{N} \sum_{i=1}^N \eta_i.
\end{equation}
The dilution fraction is then defined as $x=1-p$.

The isotropic AFM Heisenberg model corresponds to $J_a=J>0$ for
$a=x,y,z$. When LRO is present (say, along the $z$ direction), the
spin operators can be written in terms of bosonic operators using the
Holstein-Primakoff method.\cite{swt} Since the square lattice is
bipartite, it can be divided up into two square sublattices, ${\cal A}
= {\cal L}_+$ and ${\cal B} = {\cal L}_-$. Thus,
\begin{eqnarray}
\label{eq:HPA}
S_{i\, z} & = & S - a_i^\dagger a_i, \\ S_i^+ & = & \left[2S -
a_i^\dagger a_i\right]^{1/2} a_i, \\ S_i^- & = & a_i^\dagger \left[2S -
a_i^\dagger a_i\right]^{1/2},
\end{eqnarray}
with $ i\in {\cal A}$, and
\begin{eqnarray}
\label{eq:HPB}
S_{j\, z} & = & -S + b_j^\dagger b_j, \\ S_j^+ & = & b^\dagger_j
\left[2S - b^\dagger_j b_j\right]^{1/2}, \\ S_j^- & = &
\left[2S - b^\dagger_j b_j\right]^{1/2} b_j,
\end{eqnarray}
with $j\in {\cal B}$. The bosonic operators obey the usual commutation
relations, namely,
\begin{equation}
\label{eq:commut}
[a_i, a_{i^\prime}^\dagger] = \delta_{ii^\prime} \ \ {\rm and} \
\ [b_j, b_{j^\prime}^\dagger] = \delta_{jj^\prime},
\end{equation}
with all other commutators equal to zero.

For the large spin case ($S \gg 1$) or when the number of spin waves
is small ($n_i = \langle a_i\, a^\dagger_i\rangle$, $n_j = \langle
b_j\, b^\dagger_j\rangle$ $\ll S$), we can expand the square root in
Eqs. (\ref{eq:HPA}) and (\ref{eq:HPB}) in powers of $n_i$ and $n_j$,
keeping only linear terms. That allows us to write the approximate
bilinear bosonic Hamiltonian
\begin{eqnarray}
\label{eq:hamilton}
H & \approx & - JS^2 \sum_{\langle ij\rangle} \eta_i \eta_j + JS
\sum_{\langle ij \rangle} \eta_i \eta_j \left( a_i^\dagger a_i +
b^\dagger_j b_j \right. \nonumber \\ & & \left. +\; a_i b_j +
a_i^\dagger b_j^\dagger \right).
\end{eqnarray}
The first term on the right-hand side of Eq. (\ref{eq:hamilton})
represents the classical ground-state energy of the AFM. Using the
bosonic commutation relations, we can rearrange the Hamiltonian in the
following form:
\begin{equation}
H \approx - JS(S+1) \sum_{\langle ij \rangle} \eta_i \eta_j + H_{\rm
SW},
\end{equation}
where, now, the first term on the right-hand side represents the
ground state energy in the absence of quantum fluctuations, while the
latter is described by the second term,
\begin{eqnarray} 
H_{\rm SW} & = & \frac{JS}{2} \sum_{\langle ij \rangle} \eta_i \eta_j
\left( a_i^\dagger a_i + a_i a_i^\dagger + b_j^\dagger b_j + b_j
b_j^\dagger \right. \nonumber \\ & & \left. +\; a_i b_j + b_j a_i +
a_i^\dagger b_j^\dagger + b_j^\dagger a_i^\dagger \right).
\end{eqnarray}
Hereafter we will drop the constant ground state energy term and will
only study the eigenstates of $H_{\rm SW}$. The spin-wave Hamiltonian
contains bilinear crossed terms which, separately, do not conserve
particle number. At this point, it is worth simplifying the notation
by dropping the distinction between bosonic operators living on
different sublattices and reordering the summation over sites,
\begin{eqnarray}
\label{eq:HSW}
H_{\rm SW} & = & \frac{JS}{2} \sum_{i,j=1}^N \left[ K_{ij} \left(
a^\dagger_i a_j + a_i a^\dagger_j \right)
 \right. \nonumber \\ & & \left. 
+ \Delta_{ij} \left( a_i a_j + a_i^\dagger a_j^\dagger
\right)\right],
\end{eqnarray}
where both indexes in the sum run over all sites in the lattice. The
matrices $K$ and $\Delta$ are defined as
\begin{equation}
K_{ij} = \delta_{ij}\, \eta_i \sum_{\langle il\rangle} \eta_l
\end{equation}
(the sum run over all nearest-neighbor sites to $i$) and
\begin{eqnarray}
\Delta_{ij} = \left\{ \begin{array}{ll} \eta_i \eta_j & {\rm for}\
i,j \ {\rm nearest\ neighbors} \\ 0, & {\rm otherwise}. \end{array}
\right.
\end{eqnarray}
Notice that both matrices $K$ and $\Delta$ are real and symmetric.

%%%%%%%%%%%%%%%%%%%%%%%%%%%%%%%%%%%%%%%%%%%%%%%%%%%%%%%%%%%%%%%%%%%%%%%%%
\subsection{Bogoliubov transformation}

It is possible to diagonalize the spin-wave Hamiltonian through an
operator transformation of the Bogoliubov type,
\begin{eqnarray}
\label{eq:transform1}
a_i & = & \sum_n ( u_{in}\, \alpha_n + v_{in}\, \alpha^\dagger_n), \\
\label{eq:transform2}
a_i^\dagger & = & \sum_n ( v^\ast_{in}\, \alpha_n + u^\ast_{in}\,
\alpha^\dagger_n),
\end{eqnarray}
or, in matrix notation,
\begin{equation}
\label{eq:bogolubov}
\left( \begin{array}{c} \bar{a} \\ \bar{a}^\ast \end{array} \right) =
\left( \begin{array}{cc} U & V \\ V^\ast & U^\ast \end{array} \right)\
\left( \begin{array}{c} \bar{\alpha} \\ \bar{\alpha}^\ast \end{array}
\right),
\end{equation}
where the column vectors $\bar{a}$ and $\bar{\alpha}$ contain the
operators $a_i$ and $\alpha_n$, respectively, while the $N\times N$
matrices $U$ and $V$ contain the coefficients $\{u_{in}\}$ and
$\{v_{in}\}$, respectively, with $i,n=1,\ldots,N$. Assuming that the
new operators also obey the canonical commutation relations,
\begin{equation}
[\alpha_n,\alpha_m^\dagger] = \delta_{nm} \ \ {\rm and} \ \ 
[\alpha_n,\alpha_m] = [\alpha_n^\dagger,\alpha_m^\dagger] = 0,
\end{equation}
one arrives at the following constraints for the
transformation coefficients:
\begin{equation}
\label{eq:ortho1}
\sum_{n=1}^N ( u_{in}^\ast\, u_{jn} - v_{in}\, v_{jn}^\ast) =
\delta_{ij}
\end{equation}
and
\begin{equation}
\label{eq:ortho2}
\sum_{n=1}^N ( u_{in}\, v_{jn} - v_{in}\, u_{jn} ) = 0.
\end{equation}
In matrix notation,
\begin{equation}
\label{eq:orthon1}
U\, U^\dagger - V\, V^\dagger = I_N
\end{equation}
and
\begin{equation}
\label{eq:orthon2}
U\, V^T - V\, U^T = 0,
\end{equation}
where $I_N$ is the $N\times N$ unit matrix. These relations can be put
into a more compact form by defining the matrices
\begin{equation}
T = \left( \begin{array}{cc} U & V \\ V^\ast & U^\ast \end{array}
\right)
\end{equation}
and
\begin{equation}
\Sigma = \left(
\begin{array}{cc} I_N & 0 \\ 0 & - I_N \end{array} \right).
\end{equation}
Thus, Eqs. (\ref{eq:orthon1}) and (\ref{eq:orthon2}) become one,
\begin{equation}
T\; \Sigma\; T^\dagger = \Sigma.
\end{equation}
Since $\Sigma^2 = I_{2N}$, we find, after a simple algebra, that
\begin{equation}
\label{eq:orthoT}
T^\dagger \Sigma\; T = \Sigma.
\end{equation}
As a result we have two additional (though not independent) sets of
orthogonality equations,
\begin{equation}
\label{eq:ortho3}
\sum_{i=1}^N ( u_{in}^\ast\, u_{im} - v_{in}\, v_{im}^\ast) =
\delta_{nm}
\end{equation}
and
\begin{equation}
\label{eq:ortho4}
\sum_{i=1}^N ( u_{in}^\ast\, v_{im} - v_{in}\, u_{im}^\ast) = 0.
\end{equation}
%

%%%%%%%%%%%%%%%%%%%%%%%%%%%%%%%%%%%%%%%%%%%%%%%%%%%%%%%%%%%%%%%%%%%%%%%%%
\subsection{Non-Hermitian eigenvalue problem (Ref. \onlinecite{blaizotripka})}
\label{sec:nonhermitian}

The transformation defined by Eq. (\ref{eq:bogolubov}) allows us to
diagonalize the spin-wave Hamiltonian in terms of the new bosonic
operators. For that purpose, we choose $T$ such that
\begin{equation}
\label{eq:diagonal}
T^\dagger \left( \begin{array}{cc} K & \Delta \\ \Delta & K
\end{array} \right)\, T = \left( \begin{array}{cc} \Omega_+ & 0 \\ 0 &
\Omega_- \end{array} \right), 
\end{equation}
where $\Omega_\pm$ are diagonal matrices containing the
eigenfrequencies: $[\Omega_\pm]_{nn} = \omega_n^{(\pm)}$ for
$n=1,\ldots,N$. The eigenvalue problem defined by
Eq. (\ref{eq:diagonal}) can be further simplified. Recalling
Eq. (\ref{eq:orthoT}), we have that
\begin{equation}
\left( \begin{array}{cc} K & \Delta \\ \Delta & K \end{array}
\right)\, T = \Sigma\, T\, \Sigma \left( \begin{array}{cc} \Omega_+ &
0 \\ 0 & \Omega_- \end{array} \right).
\end{equation}
In fact, it is not difficult to prove that eigenfrequency matrices
obey the relation $\Omega_+ = \Omega_-^\ast = \Omega$, with $\Omega$
being a diagonal matrix with real entries only, as one physically
expects. As a result,
\begin{equation}
\left( \begin{array}{cc} K & \Delta \\ \Delta & K \end{array}
\right)\, \left( \begin{array}{cc} U & V \\ V^\ast & U^\ast
\end{array} \right) = \left( \begin{array}{cc} U & -V \\
-V^\ast & U^\ast \end{array} \right)\, \left( \begin{array}{cc} \Omega
 & 0 \\ 0 & \Omega \end{array} \right).
\end{equation}
We can break up this $2N\times 2N$ matrix equation into two coupled
$N\times N$ matrix equations,
\begin{eqnarray}
K\, U + \Delta\, V^\ast & = & U\, \Omega, \nonumber \\
\Delta\, U + K\, V^\ast & = & - V^\ast\, \Omega,
\end{eqnarray}
or, alternatively, writing explicitly the matrix elements,
\begin{eqnarray}
\label{eq:eigen1}
\sum_j \left[ K_{ij}\, u_{jn} + \Delta_{ij}\, v_{jn}^\ast \right] 
& = & \omega_n\, u_{in}, \\ 
\label{eq:eigen2}
\sum_j \left[ \Delta_{ij}\, u_{jn} + K_{ij}\, v_{jn}^\ast \right] 
& = & - \omega_n\, v_{in}^\ast,
\end{eqnarray}
for all $n$ and $i$. Thus, for a given eigenstate $n$, we can define
an eigenvalue matrix equation in the usual form, namely,
\begin{equation}
\label{eq:nohermeigen}
\left( \begin{array}{cc} K & \Delta \\ -\Delta & -K \end{array}
\right)\, \left( \begin{array}{c} u_n \\ v_n^\ast \end{array} \right) =
\omega_n\, \left( \begin{array}{c} u_n \\ v_n^\ast \end{array} \right).
\end{equation}
(Notice that each $u_n$ and $v_n$ is now a column vector with
components running through all $i=1,\ldots,N$ lattice sites.) The $2N
\times 2N$ matrix shown in Eq. (\ref{eq:nohermeigen}) is clearly
non-Hermitian, but its eigenvalues are all real. Notice also that if
$\omega_n$ is an eigenvalue with corresponding eigenvector
$(u_n,v_n^\ast)$, then $-\omega_n$ is also an eigenvalue, but with
$(v_n^\ast,u_n)$ as the corresponding eigenvector. Thus, despite the
fact that the non-Hermitian matrix provides $2N$ eigenvalues
(eigenfrequencies), we should only keep those $N$ that are positive
and whose corresponding coefficients $u_{in}$ and $v_{in}$ satisfy
Eqs. (\ref{eq:ortho1}), (\ref{eq:ortho2}), (\ref{eq:ortho3}), and
(\ref{eq:ortho4}).

The non-Hermitian matrix in Eq. (\ref{eq:nohermeigen}) contains only
integer elements: 0, 1, 2, or 4 in the diagonal (corresponding to
$K_{ii}$, i.e., the number of nearest neighbors to site $i$) and 0 or
1 in the off-diagonal components (corresponding to $\Delta_{ij}$,
i.e., 1 when $i$ and $j$ are nearest neighbors and zero otherwise). It
is strongly sparse, although without any particularly simple pattern
due to the presence of dilution disorder.

It is easy to verify that there are at least two zero modes in
Eq. (\ref{eq:nohermeigen}), i.e., two distinct nontrivial solutions
with zero eigenvalue:
\begin{equation}
u_{i0}^{(a)} = 1, \qquad v_{i0}^{(a)} = -1,
\end{equation}
for all $i=1,\ldots, N$, and
\begin{equation}
u_{i0}^{(b)} = v_{i0}^{(b)} = \left\{ \begin{array}{lr} 1, & i
\in {\cal A}, \\ -1, & i \in {\cal B}. \end{array} \right.
\end{equation}
(In order to prove that these are indeed eigenstates, notice that
$\sum_{j=1}^N \Delta_{ij} = K_{ii}$.) These two zero modes do not obey
the orthogonality relation of Eq. (\ref{eq:ortho3}); they have zero
hyperbolic norm instead.

%%%%%%%%%%%%%%%%%%%%%%%%%%%%%%%%%%%%%%%%%%%%%%%%%%%%%%%%%%%%%%%%%%%%%%%%%
\subsection{Average magnetization per site}

The total staggered magnetization can be written in terms of the
expectation value of the spin-wave number operator:
\begin{eqnarray}
M_z^{\rm stagg} & = & \left\langle \sum_{i\in {\cal A}} S_{i \, z} -
\sum_{j\in {\cal B}} S_{j \, z} \right\rangle \nonumber \\ & = & NS -
\sum_{i=1}^N \langle a_i^\dagger a_i \rangle,
\end{eqnarray}
where we have assumed that the sublattices contain the same number of
sites: $N_A = N_B = N/2$. As a result, the average staggered spin per
site along the $z$ direction can be written as
\begin{equation}
m_z = \frac{M_z^{\rm stagg}}{N} = S - \delta m_z,
\end{equation}
with
\begin{equation}
\delta m_z = \frac{1}{N} \sum_{i=1}^N \delta m_i^z,
\end{equation}
and
\begin{equation}
\delta m_i^z = \langle a_i^\dagger a_i
\rangle.
\end{equation}
Notice that $\delta m_z$ describes the spin-wave correction to the
average staggered magnetization (always a reduction).

In order to express $\delta m_z$ in terms of the coefficients $v_{in}$
and $u_{in}$, we use Eqs. (\ref{eq:transform1}) and
(\ref{eq:transform2}) to first write the site magnetization at zero
temperature in terms of eigenmodes. Upon taking the ground-state
expectation value, we have to recall that the vacuum contains zero
eigenmodes. Hence,
\begin{equation}
\langle \alpha_n \alpha_m \rangle = \langle \alpha_n^\dagger
 \alpha_m^\dagger \rangle = \langle \alpha_n^\dagger \alpha_m \rangle
 = 0,
\end{equation}
while
\begin{equation}
\langle \alpha_n \alpha_m^\dagger \rangle = \delta_{nm}.
\end{equation}
As a result,
\begin{equation}
\delta m_i^z = \sum_{n=2}^{N} |v_{in}|^2,
\end{equation}
where the sum runs only through eigenmodes with {\it positive}
frequency (the zero modes have been subtracted).

%%%%%%%%%%%%%%%%%%%%%%%%%%%%%%%%%%%%%%%%%%%%%%%%%%%%%%%%%%%%%%%%%%%%%%%%%%
\subsection{Reduction to a $N\times N$ non-Hermitian 
eigenvalue problem}
\label{sec:simplify}

It is possible to rewrite the $2N \times 2N$ eigenvalue problem as two
coupled eigenproblems, each one of order $N\times N$ instead. The
non-Hermitian character of the matrices involved does not
change. However, the amount of work for numerical computations
decreases by a factor of 4 [recall that diagonalizing a $N\times N$
requires $O(N^3)$ operations].

We begin by summing and subtracting Eqs. (\ref{eq:eigen1}) and
(\ref{eq:eigen2}), obtaining
\begin{equation}
\label{eq:1N1}
\sum_{j=1}^N (K_{ij} + \Delta_{ij})(u_{jn} - v_{jn}) = \omega_n 
(u_{in} + v_{in})
\end{equation}
and
\begin{equation}
\label{eq:1N2}
\sum_{j=1}^N (K_{ij} - \Delta_{ij})(u_{jn} + v_{jn}) = \omega_n 
(u_{in} - v_{in}).
\end{equation}
Multiplying these equations by $K-\Delta$ and $K+\Delta$, we find the
following eigenvalue equations after a simple manipulation:
\begin{equation}
\label{eq:Ndec1} 
\sum_{j=1}^N [(K - \Delta)(K + \Delta)]_{ij} (u_{jn} - v_{jn}) = 
\omega_n^2 (u_{in} - v_{in}) 
\end{equation}
and
\begin{equation}
\label{eq:Ndec2}
\sum_{j=1}^N [(K + \Delta)(K - \Delta)]_{ij} (u_{jn} + v_{jn}) = 
\omega_n^2 (u_{in} + v_{in}).
\end{equation}
Although these equations are in principle decoupled, for the purpose
of finding the local magnetization they are not so, since we are
interested in finding mainly $v$ (and not $u+v$ or $u-v$ alone). We
will come back to this point later. Equations (\ref{eq:Ndec1}) and
(\ref{eq:Ndec2}) can also be presented in more revealing form, namely
(here we will drop indices to shorten the notation),
\begin{equation}
\label{eq:newproblem}
\left( K^2 - \Delta^2 \pm [\Delta,K] \right) \phi^{(\pm)} = \lambda\, 
\phi^{(\pm)},
\end{equation}
where $\phi^{(\pm)} = u \pm v$ and $\lambda = \omega^2$. At this point
it is interesting to notice that the nonzero commutator is the cause
of non-Hermiticity in the eigenvalue problem. Had it been zero, the
problem would be become real and symmetric. In fact, it is not
difficult to show that
\begin{equation}
[\Delta,K]_{ij} = \Delta_{ij}(K_{ii} - K_{jj}).
\end{equation}
Thus, it is only when all sites have the same number or nearest
neighbors, i.e., when no dilution is present, that the problem becomes
real and symmetric (and can therefore be solved analytically by a
Fourier transform). Dilution always makes the eigenproblem
non-Hermitian, although with real eigenvalues [even if we did not know
the origin of Eq. (\ref{eq:newproblem}), it would be easy to prove
that all eigenvalues $\lambda \ge 0$].

Let us call $M^{(\pm)} = K^2 - \Delta^2 \pm [\Delta,K]$. An important
feature of these matrices is that the left eigenvector $\phi^{(\pm)}$
of $M^{(\pm)}$ is the right eigenvector of $M^{(\mp)}$.  Thus, if we
use an algorithm that is capable of finding both the right and left
eigenvectors of a non-Hermitian matrix, we only need to solve the
problem for $M^{(+)}$, for instance. In this case, we may say that the
$2N\times 2N$ problem has really been reduced to $N\times N$.

In terms of the eigenvectors $\phi^{(\pm)}$, the orthogonality
relations of Eqs. (\ref{eq:ortho3}) and (\ref{eq:ortho4}) now read
\begin{equation}
\label{eq:newsumrule1} 
\sum_{i=1}^N \left[ \phi_{in}^{(+)\; \ast} \phi_{im}^{(-)} + 
\phi_{in}^{(-)\; \ast} \phi_{im}^{(+)} \right] = 2\, \delta_{nm} 
\end{equation}
and
\begin{equation}
\label{eq:newsumrule2} 
\sum_{i=1}^N \left[ \phi_{in}^{(-)\; \ast} \phi_{im}^{(+)} - 
\phi_{in}^{(+)\; \ast} \phi_{im}^{(-)} \right] = 0,
\end{equation}
respectively. Moreover, using these relations and the definition of
$\phi^{(\pm)}$, it is straightforward to show that the average site
magnetization can be written as
\begin{equation}
\delta m_z = \frac{1}{4N} \sum_{n=2}^N \sum_{i=1}^N \left[ \left| 
\phi_{in}^{(+)} \right|^2 + \left| \phi_{in}^{(-)} \right|^2 \right]
- \frac{1}{2}.
\end{equation}

One issue that appears when diagonalizing the problem through solving
Eq. (\ref{eq:newproblem}) is that each eigenstate $\lambda_n$ may have
an eigenvector corresponding to any linear combination of the $u_n$
and $v_n$ vectors, and not just that $u_n \pm v_n$ (that is because
each $\lambda_n$ corresponds to at least two eigenfrequencies, namely
$\pm \omega$, with $\omega_n =\sqrt{\lambda_n}$). Provided that there
are no other degeneracies, one can sort out which combination is
generated by noticing the following. Suppose that
\begin{equation}
\label{eq:linearcomb}
\phi^{(+)} = c_+\, ( u - v) \qquad {\rm and} \qquad \phi^{(-)} =
c_-\, ( u + v),
\end{equation}
then, it is easy to see that the normalization conditions for both
$\phi^{(\pm)}$ and $u,v$ imply $c_+c_-=1$. We can then use
Eqs. (\ref{eq:1N1}) and (\ref{eq:1N2}) to find that
\begin{equation}
(K - \Delta)\, \phi^{(+)} = \frac{\omega}{c_+^2} \, \phi^{(-)}
\end{equation}
and
\begin{equation}
(K + \Delta)\, \phi^{(-)} = c_+^2\, \omega\, \phi^{(+)}.
\end{equation}
These equations provide a way of determining the coefficient $c_+$
(and thus the actual mixing of degenerate eigenvectors). For instance,
for a given eigenstate,
\begin{equation}
\label{eq:coeff}
c_+^2 = \frac{1}{\sqrt{\lambda}} \frac{\sum_{i,j=1}^N \phi^{(+)\,}_i\,
(K+\Delta)_{ij}\, \phi_j^{(-)}}{\sum_{i=1} \phi^{(+)}_i\,
\phi^{(-)}_i}.
\end{equation}
Once the $c_+$ coefficient has been determined, it is straightforward
to determine the $u_n$ and $v_n$ vectors corresponding to a given
(positive), nondegenerate eigenfrequency $\omega_n$ in a unique
way. If additional degeneracy occurs, then one needs to introduce more
coefficients (and consider combinations of all degenerate
eigenvectors) in Eq. (\ref{eq:linearcomb}).

%%%%%%%%%%%%%%%%%%%%%%%%%%%%%%%%%%%%%%%%%%%%%%%%%%%%%%%%%%%%%%%%%%%%%%%%%
%%%%%%%%%%%%%%%%%%%%%%%%%%%%%%%%%%%%%%%%%%%%%%%%%%%%%%%%%%%%%%%%%%%%%%%%%
\section{Numerical solution of the non-Hermitian eigenproblem}
\label{sec:numericalmethod}

The numerical solution of the eigenproblem represented by
Eq. (\ref{eq:newproblem}) requires the full diagonalization of at
least one real nonsymmetric matrix. However, before that, we need to
generate the random dilution on a square lattice and set the
appropriate boundary conditions. Another important point is that we
can simplify the diagonalization by breaking the matrix into diagonal
blocks, each one related to a single disconnected cluster. The
diagonalizations can then be carried out separately on each block (for
each disconnected cluster). Thus, the first task is to reorganize the
matrices $K$ and $\Delta$ following a hierarchy of disconnected
cluster sizes. That involves only searching and sorting sites on the
lattice (without any arithmetic or algebraic manipulation). Moreover,
since we are interested only in what happens within the largest
connected cluster (the only one relevant in the thermodynamic limit
and below the percolation threshold),\cite{stauffer94} we can
concentrate our numerical effort into the diagonalization of the
matrix block corresponding to that cluster alone.

The first step is to create a square lattice of size $N = L\times L$
($L$ being the lateral size of the lattice) with periodic boundary
conditions in both directions. In order to have $N_A = N_B$, we choose
$L$ to be an even number. We fix the number of holes as the integer
part of $(1-p)N$ and randomly distribute them over the lattice with
uniform probability.

The second step is to identify all connected clusters that exist in
the lattice for a given realization of dilution. Since most lattices
we work with are quite dense (relatively few holes), we begin by
finding all sites that belong to the cluster whose sites are nearest
to one of the corners of the lattice. Once all sites in that cluster
are found, they are subtracted from the lattice and the search begins
again for another cluster. The process stops when all sites have been
visited and the whole lattice is empty. The process of identifying
sites for a particular cluster is the following. Starting from a fixed
site $i$ (the cluster seed), we check whether its four neighbors are
occupied or empty. The occupied ones get the same tag number as the
first site visited. Then we move on to the next side, $i+1$, and
repeat the procedure. We continue until we reach the $N$th site.

Along with identifying all sites belonging to each cluster, we also
count then. That allows us to identify immediately the largest
connected cluster in the lattice, whose number of sites we call
$N_c$. We set a conversion table where the sequential number
identifying a site in the largest cluster is associated to its
coordinate in the original lattice. That allows us to later retrace
the components of the eigenvectors of this cluster to their locations
in the $L\times L$ lattice.

The process of identifying clusters is carried out under hard wall
boundary conditions. In fact, it is only after the largest cluster is
found that we force periodic boundary conditions. This is the third
step. For that purpose, we sweep the bottom and top rows, as well as
the right and left columns of the square lattice, and check whether
these sites are neighbors of sites belonging to the largest cluster
once periodic boundary conditions are assumed. It turns out that it is
easier and faster to do that than to search and classify clusters
directly from a lattice with periodic boundary conditions.

The fourth step consists of storing the information necessary to
assemble the non-Hermitian matrices of the type shown in
Eq. (\ref{eq:nohermeigen}) for the largest cluster only. The algorithm
is quite fast and allows one to generate and find the largest
connected cluster in lattices as large as $L=100$ in less than a
second.

The information generated in the process of identifying the largest
cluster and its structure is fed into a second routine. There, one
assembles both the non-Hermitian matrices of
Eqs. (\ref{eq:nohermeigen}) and (\ref{eq:newproblem}). We have checked
that the solution of both the $2N_c \times 2N_c$ and $N_c \times N_c$
problems provide identical solutions up to several digits for a
particular realization of the dilution problem at various lattice
sizes and dilution fractions. However, only the $N_c \times N_c$
problem was used to generate the data presented here.

It is important to point out that there exists an alternative
formulation of the problem defined by Eq. (\ref{eq:HSW}), using
generalized position and momentum operators (see Appendix
\ref{sec:appendixA}). In this formulation one can derive a sequence of
transformations that permits the calculation of eigenvalues and
eigenvectors of the system Hamiltonian through the diagonalization of
{\it Hermitian} matrices alone. However, from the computational point
of view there is no substantial advantage of this approach with
respect to the non-Hermitian one.

Since the solution of the non-Hermitian eigenvalue problem is less
standard than the Hermitian case, we provide a description of the
method in Appendix \ref{sec:appendixB}

%%%%%%%%%%%%%%%%%%%%%%%%%%%%%%%%%%%%%%%%%%%%%%%%%%%%%%%%%%%%%%%%%%%%%%%%%
%%%%%%%%%%%%%%%%%%%%%%%%%%%%%%%%%%%%%%%%%%%%%%%%%%%%%%%%%%%%%%%%%%%%%%%%%
\section{Results}
\label{sec:numericalresults}

We have generated lattices with sizes ranging from $L=12$ to $36$ and
dilution fractions going from $x=0$ to $x_c$. The number of
realizations for a given size and dilution fraction varied between 500
(nearly clean case) to 1000 (at the classical percolation threshold).
The results of the numerical diagonalizations are described below.

%%%%%%%%%%%%%%%%%%%%%%%%%%%%%%%%%%%%%%%%%%%%%%%%%%%%%%%%%%%%%%%%%%%%%%%%%
\subsection{Density of states}

%%%%%%%%%%%%
%
\begin{figure}
\includegraphics*[width=7.8cm]{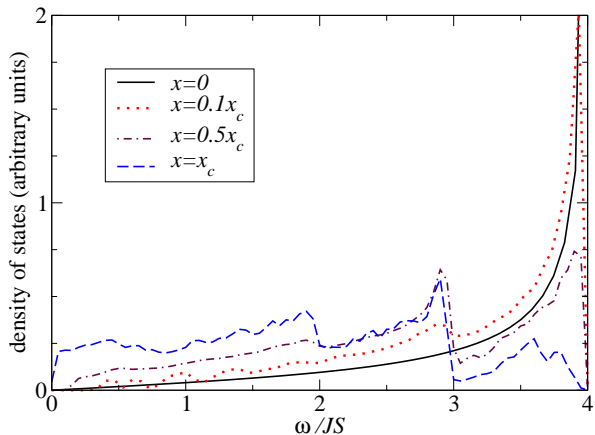}
\caption{Average density of states as a function of energy for four
different site dilution fractions: $x=0$ (solid line), $x=0.1\, x_c$
(dotted line), $x=0.5\, x_c$ (dotted-dashed line), and $x=x_c$ (dashed
line). Notice the two structures just below $\omega/JS = 2$ and
$3$. The latter is already visible at $x=0.1\, x_c$, while the former
becomes prominent only for $x\agt 0.4\, x_c$. The data was obtained
from 500 to 1000 realizations of a $36\times 36$ lattice. }
\label{fig:dos}
\end{figure}  
%
%%%%%%%%%%%%

The ensemble averaged density of eigenstates as a function of
frequency for $L=36$ is presented in Fig. \ref{fig:dos} for several
dilution fractions and compared to the well-known result for the clean
case.\cite{auerbach94} For a small dilution, there is little departure
from the clean case, although a small structure is already visible at
around $\omega/JS=3$. As the dilution increases, a peak and an edge
develop at around this frequency. Notice that the overall trend is a
decrease in the number of high-frequency modes, with the proportional
increase in the number of low-frequency ones. Close to the percolation
threshold, another structure appears at around $\omega/JS=2$. Thus, we
see that the effect of dilution is to shift spectral weight from high
to low frequencies in a nonuniform way. This tendency had also been
observed in Ref. \onlinecite{sasha} for small dilution.

\begin{figure}
\includegraphics[width=6cm]{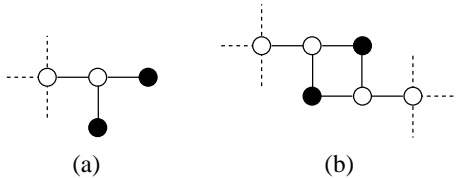}
\caption{The two dangling structures that occur in the $\omega/J=1$
(a) and $2$ (b) eigenstates. The filled circles indicate $u_{in} \neq
0$, while the empty circles have $u_{in} = 0$. All sites in these
states have $v_{in} = 0$, thus they do not contribute to the staggered
magnetization.}
\label{fig:pendura}
\end{figure}

Two additional very sharp peaks (not shown in Fig. \ref{fig:dos}) also
exist in the density of states as the dilution increases. They occur
at frequencies $\omega/JS = 1,2$ and correspond to configurations
where $v_{in} = 0$ for all sites in the cluster, while $u_{in} = 0$
for all but two sites. Their typical spatial structures are shown in
Fig. \ref{fig:pendura}. Since $v_{in} = 0$ for all sites, these states
do not contribute to the quantum corrections to the staggered
magnetization.

\begin{figure}
\includegraphics*[width=7.8cm]{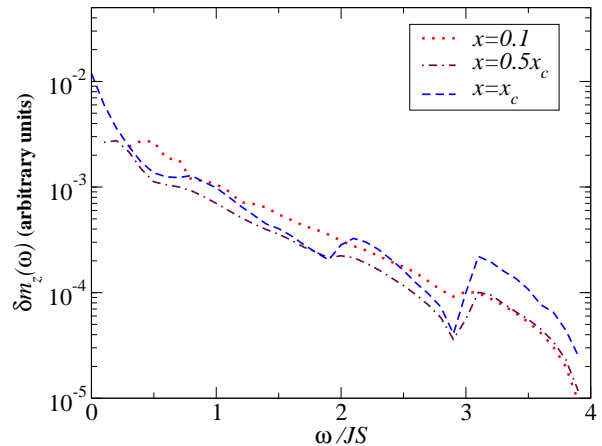}
\caption{Staggered magnetization per unit of magnetic site as a
function of energy, $\delta m_z(\omega)$, under the same conditions of
Fig. \ref{fig:dos}.}
\label{fig:mag}
\end{figure}

For the clean case, it is simple to verify (based on the exact
diagonalization of the problem) that the low frequency modes provide
the largest contribution to $\delta m_z$. For the dilute lattice, the
same is true, as can be seen in Fig. \ref{fig:mag}, where we have
plotted
\begin{equation}
\delta m_z(\omega) = \frac{1}{N} \frac{\left\langle \sum_{i=1}^N
\sum_{n=2}^N \delta m_i^z \, \delta(\omega - \omega_n) \right\rangle}
{\left\langle \sum_{n=2}^N \delta (\omega - \omega_n) \right\rangle}.
\end{equation}
As a result, we see that the transference of eigenmodes from high to
low frequencies is the mechanism by which quantum fluctuations are
enhanced as the dilution increases.

%%%%%%%%%%%%
%
%
%%%%%%%%%%%%%

%%%%%%%%%%%%
%
%
%%%%%%%%%%%%

%%%%%%%%%%%%%%%%%%%%%%%%%%%%%%%%%%%%%%%%%%%%%%%%%%%%%%%%%%%%%%%%%%%%%%%%%
\subsection{Inverse participation ratio}

The nature of the eigenstates also changes as the dilution
increases. The best way to characterize the nature of the states is
through the return probability, that is, the probability that after
some very long time a particle, moving in the percolating lattice,
will return to its originating point. The return probability can be
expressed in terms of the inverse participation ratio
(IPR).\cite{wegner80} Here, we use a definition of the IPR involving
the eigenvector component related to the quantum fluctuation
corrections to the magnetization, namely,
\begin{equation}
\label{eq:invpartw}
I(\omega) = \frac{\sum_{n=2}^N \delta(\omega - \omega_n) I_n}
{\sum_{n=2}^N \delta(\omega - \omega_n)},
\end{equation}
where
\begin{equation}
\label{eq:invpartn}
I_n = \frac{\sum_{i=1}^{N_c} v_{in}^4}{\left(\sum_{i=1}^{N_c}
v_{in}^2\right)^2}.
\end{equation}

\begin{figure}  
\includegraphics[width=7.8cm]{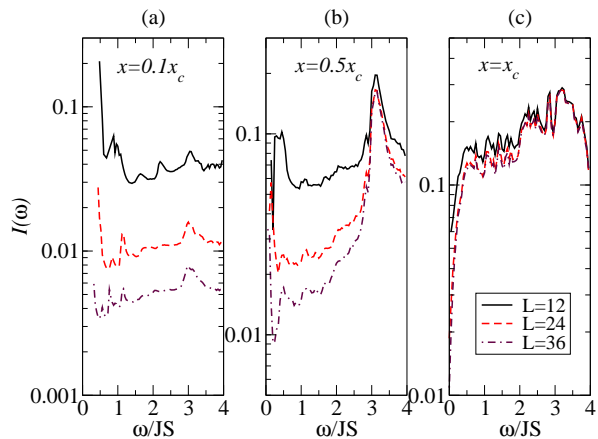}
\caption{Average inverse participation ratio, as defined in
Eq. (\ref{eq:invpartw}), as a function of energy for different lattice
sizes and dilution fractions: (a) $x=0.1\, x_c$, (b) $x=0.5\, x_c$,
and (c) $x=x_c$. As the dilution increases, states become more
localized, beginning with those located in the high-energy part of the
spectrum. Each curve shown corresponds to an average over 500 to 1000
realizations.}
\label{fig:3}
\end{figure}

In Fig. \ref{fig:3} we show the IPR as a function of energy for three
lattice sizes. According to its definition, the IPR for extended
states decrease as the system size increases, while for localized
states the IPR is insensitive to any size variation. These trends are
clearly visible in Fig. \ref{fig:3}, namely, states are mostly
extended when dilution is small and tend to localize as one gets
closer to the percolation threshold. For intermediate dilution [Fig.
\ref{fig:3}(c)], we see that the states close to $\omega/JS=3$ are
strongly localized while the remaining states are quite extended. As
expected, the low frequency states tend to remain extended up to
strong dilution.

%%%%%%%%%%%%%
%
%
%%%%%%%%%%%%%

%%%%%%%%%%%%%%%%%%%%%%%%%%%%%%%%%%%%%%%%%%%%%%%%%%%%%%%%%%%%%%%%%%%%%%%%%
\subsection{Average magnetization}
\label{sec:magnetization}

In the thermodynamic limit, the magnitude of the staggered
magnetization per unit of lattice site $m_z$ can be written as
\begin{equation}
\label{eq:clustercontr}
m_z = \left( \frac{N_c}{N_m} \right) \left( S - \delta m_z \right) ,
\end{equation}
where $N_m$ is the total number of occupied sites in the lattice ($N_m
= p N$). While the first factor on the right-hand side of
Eq. (\ref{eq:clustercontr}) is purely classical, the second factor is
purely quantum, namely, it measures how quantum fluctuations reduce
magnetic order. Thus, since $N_c$ vanishes at $x_c$, a quantum
critical point below $x_c$ can only exist if, for some $x^\ast < x_c$,
we find $\delta m_z = S$. If, on the contrary, $\delta m_z < S$ at
$x=x_c$, then the order is only lost at the percolation threshold and
the transition is essentially classical. It is important to have in
mind that the linear spin-wave approximation is well defined only when
$\delta m_i^z \ll S$, for all $i = 1,\ldots,N$. When $\delta m_i^z
\approx S$ at a large number of sites, the approximation is not
necessarily quantitatively correct.

\begin{figure}  
\includegraphics[width=8cm]{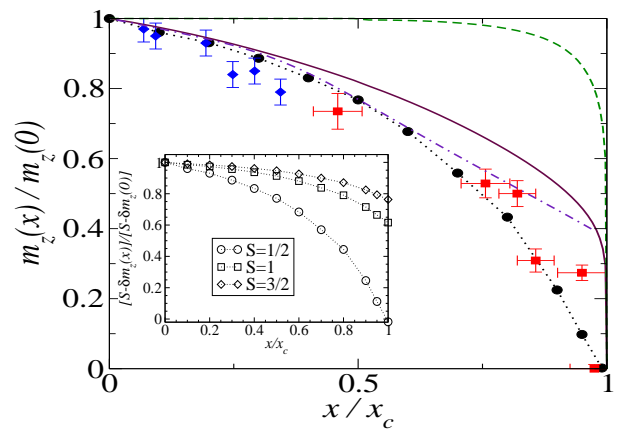}
\caption{Average staggered magnetization per unit of magnetic
site. Results of the $S=\frac{1}{2}$ SWT simulations after finite-size
scaling (circles) are compared with neutron scattering data
(squares),\cite{vajk02science} NQR data (diamonds),\cite{corti95} and
the fit to the QMC data from Ref. \onlinecite{sandvik02prb} (solid
line). Also shown is the occupation fraction of the largest connected
cluster, $N_c/N_m$, (dashed line) and the analytical result from the
calculation of Ref. \onlinecite{sasha} (dashed-dotted line). Inset:
the quantum fluctuation contribution to the staggered magnetization
for different values of $S$.}
\label{fig:1}
\end{figure}

Figure \ref{fig:1} shows the average staggered magnetization per unit
of magnetic site as a function of dilution fraction, $m_z(x)$, when
$S=1/2$. The points were obtained after finite-size scaling the
ensemble averaged data taken from 12 different lattices sizes. As a
consistency check, we have also calculated the staggered magnetization
for the clean case ($x=0$, no ensemble average) with the same
numerical procedure. We have found that $m_z(0) \approx 0.303$,
consistent with values obtained by other methods.\cite{cleansystemQF}
Thus, at least at low dilution, the spin-wave approximation is quite
accurate.

%%%%%%%%%%%%%%
%
%
%%%%%%%%%%%%%%

For comparison, we also show experimental data obtained for LCMO from
both neutron scattering \cite{vajk02science} and nuclear quadrupole
resonance (NQR),\cite{corti95} as well as the result of QMC
simulations of the dilute Heisenberg AFM in a square
lattice.\cite{sandvik02prb} One can see that our simulations, based on
the spin-wave approximation, capture the main features of the
experimental data, namely, a progressive decrease of the staggered
magnetization up to the classical percolation threshold. At a dilution
fraction very close to $x_c$, our simulations indicate that the
staggered magnetization should vanish. The inset in Fig. {\ref{fig:1}
shows that the vanishing of the staggered magnetization occurs because
$\delta m_z$ goes to $1/2$ very close to the classical transition
point. Thus, the same effect would not arise had we used $S>1/2$. The
QMC simulations, on the other hand, predicts $\delta m_z < 1/2$ at
$x=x_c$, thus indicating that the transition is purely classical. The
relatively small number of experimental points and the large error
bars near the percolation threshold do not allow for an adequate
distinction between a classical and a quantum transition for LCMO.

The discrepancy between our result and the QMC simulations for the
staggered magnetization close to $x_c$ should be seen as an indication
that, while qualitatively correct, our approach fails quantitatively
when the order parameter magnitude is significantly reduced
locally. This is expected if we recall the assumption used in the
derivation of Eq. (\ref{eq:hamilton}). Nevertheless, the spin-wave
approximation, having access to low-lying excited states and wave
functions, allows us to understand in more detail, at least
qualitatively, how the suppression of order due to quantum
fluctuations takes place upon dilution. This is not the case for the
QMC simulations. In fact it is surprising that our calculations seem
to agree with the experimental data better than the QMC. This can be
understood by the fact that the experimental system may contain extra
oxygen atoms that introduce holes in the CuO$_2$ planes, as well as
next-nearest neighboring interactions, that frustrate the AFM state
and also introduce larger quantum fluctuations. The latter are
captured by the overestimation produced in the linear spin-wave
theory. In fact, it is known that La$_2$CuO$_4$, has a nonzero
frustrating next-nearest neighbor coupling.\cite{vajk02science} That
effectively decreases the spin per site to a value smaller than 1/2,
possibly bringing LCMO closer to a quantum critical point than the
pure $S=1/2$ Heisenberg AFM.

%%%%%%%%%%%%%%%%%%%%%%%%%%%%%%%%%%%%%%%%%%%%%%%%%%%%%%%%%%%%%%%%%%%%%%%%%
\subsection{Local fluctuations}

We now turn to the question of local fluctuations. We have so far
discussed the site-averaged demagnetization $\delta m_z$ and used the
criterion that it must be smaller than $S$ for the order to
persist. However, one could argue that some sites may have
particularly large fluctuations; if these large fluctuations take
place exactly at weak links of the largest connected cluster backbone,
then they could be responsible for earlier destruction of the
long-range order.\cite{Ali-Moore} We have numerical evidence that this
is unlikely, although the relatively small size of our lattices does
not allow us to be conclusive. In Fig. \ref{fig:colormap} we show an
intensity plot of the local quantum fluctuations $\delta m_i$ in the
largest connected cluster, very close to the percolation threshold,
for a typical realization of a $L = 32$. Notice that the largest
fluctuations tend to appear only along deadends or dangling
structures, and not in the links connecting blobs of the cluster
backbone. The same trend is seen in all realizations that we have
inspected.

%%%%%%%%%%%%
%
\begin{figure}  
\includegraphics[width=8cm]{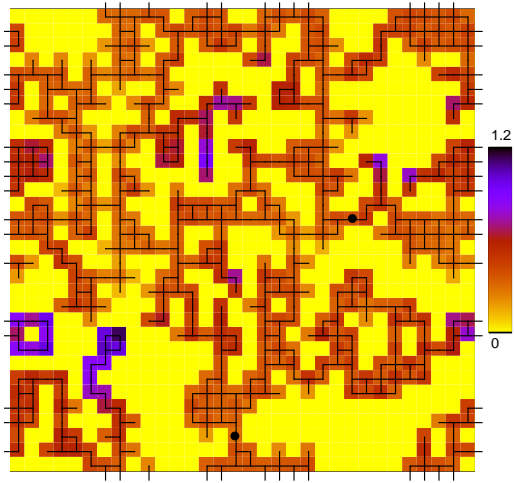}
\caption{The density plot of local demagnetization $\delta m_i^z$ for
a $32\times 32$ lattice with periodic boundary conditions at $x\approx
x_c$. Only sites belonging to the largest connected cluster are
shown. The weak links present in the cluster are indicated by
bullets.}
\label{fig:colormap}
\end{figure}  
%
%%%%%%%%%%%%

%%%%%%%%%%%%%%%%%%%%%%%%%%%%%%%%%%%%%%%%%%%%%%%%%%%%%%%%%%%%%%%%%%%%%%%%%
\subsection{Excited states}

Equation (\ref{eq:HSW}) represents a system of $N$ coupled harmonic
oscillators with $A = K + \Delta$ and $B = K - \Delta$ being the
spring constant and the inverse mass tensors, respectively (for an
alternative description of the problem in terms of position and
momentum operators, see Appendix \ref{sec:appendixA}). For the simple
model of elastic vibrations in a lattice, when $q_i = (a_i +
a_i^\dagger)/\sqrt{2}$ has the meaning of a displacement of the $i$th
atom around its equilibrium position, it is well known that the
Hamiltonian of such a system can be mapped onto a problem of diffusion
in a disordered lattice.\cite{nakayama94} Analytical results for the
related diffusive problem, as well as numerical simulations with large
systems, allow one to understand several properties of the diluted
vibrational model, such as the density of states (DOS) and the
dynamical structure factor. Perhaps one of the most distinctive
features is the existence of strongly localized excitations named
fractons.\cite{alexander82} For dilution fractions $x<x_c$, these
excitations appear in the high-frequency portion of the spectrum,
beyond a certain crossover scale $\omega_c$, while the low-frequency
part is dominated by acoustic phononlike, nearly extended,
excitations. At exactly $x=x_c$, the systems becomes a fractal and the
fractons take over the entire spectrum.\cite{nakayama94} Scaling
considerations, as well as numerical simulations, have shown that, for
a square lattice, the DOS behaves as
\begin{equation}
\rho(\omega) \propto \left\{ \begin{array}{cc} \omega, & \omega <
\omega_c \\ \omega^{1/3},& \omega > \omega_c. \end{array} \right.
\end{equation}
The crossover frequency depends critically on the dilution: $\omega_c
\propto (x_c - x)^{D}$, where $D\approx 91/48$. This is an important
result because, for the elastic vibration problem, it is also possible
to show that the quantum fluctuation corrections to the classical
order parameter obey the relation
\begin{equation}
\delta m = \frac{1}{N} \sum_{n=1}^N \frac{1}{\omega_n}
\stackrel{N\rightarrow \infty}{\longrightarrow}
\int_0^{\omega_\text{max}} d\omega\; \frac{\rho(\omega)} {{\omega}},
\end{equation}
where $\{\omega_n\}$ are the nonzero eigenvalues of the corresponding
Hamiltonian. Therefore, $\delta m$ remains finite below and at the
percolation threshold, indicating that quantum fluctuations are likely
not sufficiently enhanced by the dilution to destroy the existent
long-range order.\cite{senthil} We will get back to this point in
Sec. \ref{sec:bosons}.

In order to investigate the existence of such fractons in the dilute
2D Heisenberg AFM, as well as to clarify the nature of its low-lying
excitations, we have calculated the the dynamical structure factor,
\begin{eqnarray}
%\begin{equation}
\label{eq:sqw}
{\cal S}({\bf q},\omega)
 & = & 
\int dt\, e^{-i\omega t}
\sum_{i,j=1}^{N_m} e^{i {\bf q} \cdot ({\bf R}_i - {\bf R}_j)}
\nonumber \\ & & \times 
\left\langle S_i^+(0) S_j^-(t) + S_i^-(0)
S_j^+(t) \right\rangle .
\end{eqnarray}
%\end{equation}
%
By using Eqs. (\ref{eq:HPA}), (\ref{eq:HPB}), and
(\ref{eq:bogolubov}), we can express ${\cal S}({\bf q},\omega)$ in
terms of Fourier transformations of the site-dependent Bogoliubov
coefficients $u_{in}$ and $v_{in}$. We find
\begin{eqnarray}
{\cal S}({\bf q},\omega) & = & 2S \sum_{\omega_n\neq 0} \delta (\omega
- \omega_n) \Big[ \tilde{u}^{\cal A}_n({\bf q})\; \tilde{u}^{\cal
A}_n(-{\bf q}) \nonumber \\ & & + \tilde{v}^{\cal A}_n({\bf q})\;
\tilde{v}^{\cal A}_n(-{\bf q}) + \tilde{u}^{\cal A}_n({\bf q})\;
\tilde{v}^{\cal B}_n(-{\bf q}) \nonumber \\ & & + \tilde{v}^{\cal
A}_n({\bf q})\; \tilde{u}^{\cal B}_n(-{\bf q}) + \tilde{u}^{\cal
B}_n({\bf q})\; \tilde{v}^{\cal A}_n(-{\bf q}) \nonumber \\ & & +
\tilde{v}^{\cal B}_n({\bf q})\; \tilde{u}^{\cal A}_n(-{\bf q}) +
\tilde{v}^{\cal B}_n({\bf q})\; \tilde{v}^{\cal B}_n(-{\bf q})
\nonumber \\ & & + \tilde{u}^{\cal B}_n({\bf q})\; \tilde{u}^{\cal
B}_n(-{\bf q}) + \Big],
\end{eqnarray}
where the partial terms involving $\tilde{u}_n^{\cal A,B}$ are given
by the Fourier transformation of $u_n$, namely,
\begin{equation}
\tilde{u}_n^{\cal A,B}({\bf q}) = \sum_{i\in {\cal A,B}} u_{in}\, e^{i
{\bf q} \cdot {\bf r}_i},
\end{equation}
and analogously for $\tilde{v}_n^{\cal A,B}$. Notice that the sum over
sites runs only over one of the sublattices, ${\cal A}$ or ${\cal B}$,
depending on the particular term. Thus, only four two-dimensional
Fourier transformation are required in order to evaluate ${\cal
S}({\bf q},\omega)$.

%%%%%%%%%%%%
%
\begin{figure}  
\includegraphics[width=9cm]{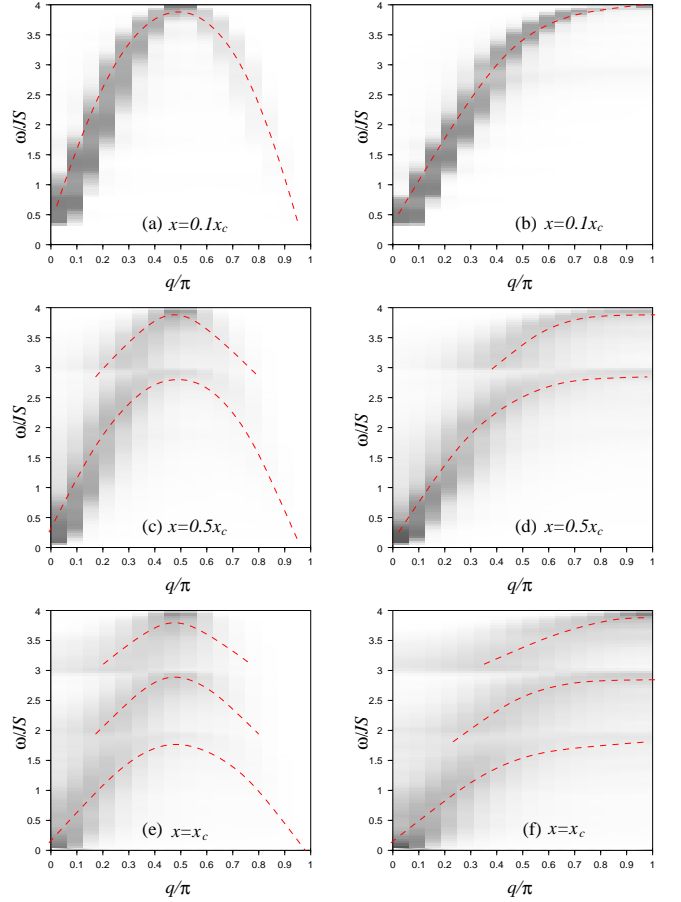}
\caption{Gray plots of the rescaled average dynamical structure factor
sliced along two distinct directions in momentum space: $q_x=q_y=q$
[(a), (c), and (e)] and $q_x=q,q_y=0$ [(b), (d), and (f)]. Plots at
different dilution fractions do not necessarily have the same
arbitrary scale for the gray intensity. The most prominent branches
are marked with dashed lines. Averaging for each plot was performed
over 50 realizations of dilution.}
\label{fig:4}
\end{figure}  
%
%%%%%%%%%%%%

We have computed numerically the Fourier transformations and
calculated the average dynamical structure factor for lattices of size
$L=32$ at several dilution fractions. Only sites within the largest
connected cluster were taken into account. Averages were performed
over 50 realizations for each case. The results are presented in the
form of intensity plots in Fig. \ref{fig:4}. To provide a better
contrast, we have rescaled $\langle {\cal S}({\bf q},\omega) \rangle$
by the function, $f(\omega) = \sum_q \langle {\cal S}({\bf q},\omega)
\rangle$. Only the data along two particular directions in momentum
space are shown, namely $q=q_x=q_y$ and $q=q_x,q_y=0$, with $0\leq q
\leq \pi$ (the lattice spacing is taken to be unit). For small
dilution, the structure factor resembles closely that of the clean
case, with some small broadening of the magnon branch due to the weak
destruction of translation invariance. However, particularly along the
$q_y=0$ direction, one can already notice a small hump at around
$\omega/JS = 3$, consistent with the peak-and-edge structure seen in
Fig. \ref{fig:dos}. This feature becomes more prominent with
increasing dilution. For dilution fractions larger than $0.6x_c$,
another hump becomes visible at around $\omega/JS = 2$, again
consistent with the feature observed in Fig. \ref{fig:dos} at the same
frequency. Close to the percolation threshold, there exist three clear
broad branches in the spectrum. While the dispersion of the
high-frequency branch at $\omega/JS > 3$ is hardly affected by the
dilution, the opposite occurs with the low-frequency one, at
$\omega/JS < 2$, where the slope (magnon velocity) decreases with
increasing dilution. We interpret the progressing breaking and bending
of the magnon branch as the system becomes more diluted to the
appearance of fractons. At $x=x_c$ the excitation spectrum has little
resemblance to that of $x=0$ and even the long wavelength part is
strongly modified. In between these two limits, there is a crossover
from a magnon-dominated to a fracton-dominated spectrum. A
three-branch structure for the spectral function in the spin-wave
approximation was also found in Ref. \onlinecite{sasha}. However, the
positioning of those branches and their relative intensity were
different from what we observed in our numerical solution. We believe
the cause of this discrepancy is the limited range of applicability of
the perturbative treatment, which is expected to be accurate only in
the weak dilution regime.

%%%%%%%%%%%%
%
\begin{figure}  
\includegraphics[width=8cm]{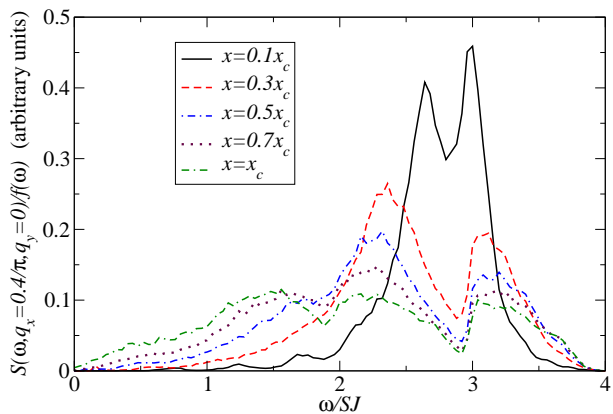}
\caption{Slice of the rescaled average dynamical structure factor for
$q_x=0.4/pi$ and $q_y=0$ for $L=32$ and different dilution fractions.}
\label{fig:5}
\end{figure}  
%
%%%%%%%%%%%%

The gradual appearance, broadening, and motion of the branches are
better represented in Fig. \ref{fig:5}, where the rescaled average
dynamical structure factor is shown as a function of frequency for a
fixed momentum $q_x = 0.4/\pi$ and $q_y = 0$. The two-peak structure
observed in our numerical data has some resemblance to the results of
inelastic neutron scattering performed by Uemura and Birgeneau for the
compound Mn$_x$Zn$_{1-x}$F$_2$,\cite{uemura87} whose magnetic
properties are described by the three-dimensional site-diluted random
Heisenberg model with $S=5/2$. These authors observed two broad peaks
which were associated with magnons (low-energy, extended) and fractons
(high-energy, localized) excitations. The relative amplitude, width,
and dispersion relations of these excitations were measured and were
found consistent with the theoretical predictions by Orbach and Yu
\cite{orbachyu} and Aharony and co-workers.\cite{aharony} The scaling
theory of the latter predicts that a two-peak spectrum appears at
momenta $q\sim \xi^{-1}$, where $\xi$ is the percolation correlation
length (i.e., the typical linear size of the disconnected clusters
when $x<x_c$). The relatively small size of the $L=32$ lattice does
not allow us to make a similar quantitative comparison between these
theories and our numerical results.

While we do observe a two-peak structure, we also see a third peak at
sufficiently large diluation, although quite weak. This additional
peak may be characteristic of square lattices; in the
body-centered-cubic Mn$_x$Zn$_{1-x}$F$_2$, there might exist a
different multipeak structure. Moreover, the large broadening and
limited energy resolution of the neutron-scattering experiments may
have made such features unobservable in the data of
Ref. \onlinecite{uemura87}. However, one aspect which seems different
between the experimental data and our numerical simulations is the way
the spectral weight is transfered between magnons and fractons as a
function of dilution. While Uemura and Birgeneau see an increase
(decrease) of high-frequency fractons (low-frequency magnons), we see
the opposite: The low-frequency portion of the spectrum becomes more
intense.

The averaging procedure recovers, to some extent, the translation
invariance broken by the dilution. This allowed us to identify the
large momentum part of the branches which, otherwise, would not be
visible. However, the low momentum part is clearly visible even when
no averaging is performed on the data (not shown). This fact, together
with the strong dispersion of the lower energy branches, suggests that
long wavelength propagating magnons are present for dilution fractions
below the percolation threshold, and lose out to fractons at
$x=x_c$. At $T=0$, these modes are responsible for the quantum
fluctuations that bring down the long range order. Their spectral
weight becomes more important as the high-frequency states become more
localized with increasing dilution. This feature is masked in
Fig. \ref{fig:4} by the frequency-dependent rescaling, but is clear
from Fig. \ref{fig:dos}. The high-frequency, high momentum modes are
much less dispersive than the low-frequency, low momentum ones. In
fact, particularly below $\omega/JS = 3$, the magnon branch is $q$
independent (i.e., broad in momentum) and thus strongly localized.

The low-frequency modes are likely not fully ballistic (coherent),
given the large broadening seen in the structure factor, but should
rather have a diffusive propagation at large scales. As to whether
they remain two dimensional or gain a lower-dimensional character, our
data is not conclusive.

%%%%%%%%%%%%%%%%%%%%%%%%%%%%%%%%%%%%%%%%%%%%%%%%%%%%%%%%%%%%%%%%%%%%%%%%%
%%%%%%%%%%%%%%%%%%%%%%%%%%%%%%%%%%%%%%%%%%%%%%%%%%%%%%%%%%%%%%%%%%%%%%%%%
\section{Upper Bound for Quantum Fluctuations in Bosonic Systems}
\label{sec:bosons}

The possibility of a classification of quantum random systems into
universality classes is an important theoretical problem with
relevance to experiments. Random matrix classification schemes
introduced by Wigner limited the classes of problems initially to the
orthogonal, unitary, and symplectic classes,\cite{Wigner} but these
were later extended to encompass the classes of chiral models
\cite{Gade} and models where the particle number is not a conserved
quantity.\cite{Zirnbauer} While most of the random matrix problems are
related to fermionic spectra, there is renewed interest in the problem
of bosonic random matrix theory.\cite{Gurarie-Chalker} Of particular
importance is its application to the problem of diluted quantum
magnets since these systems, in the limit of large spin $S$, can be
approximately described, via linear spin-wave theory, as a problem of
noninteracting bosons.\cite{swt}

One of the main differences between fermions and bosons is that, in
addition to the symmetries of the underlying Hamiltonian, one must
ensure that the bosonic spectrum is semipositive definite; this
stability condition is not an issue in fermionic systems. However, it
is automatically satisfied when the disorder is caused by site
dilution.

Let us concentrate our discussion on systems whose Hamiltonian can be
mapped onto a set of $N$ coupled harmonic oscillators of the following
kind:
\begin{equation}
\label{eq:coupledHO}
H = \frac{1}{2} \sum_{i,j=1}^N \left( q_i\, A_{ij}\, q_j + p_i\,
B_{ij}\, p_j \right).
\end{equation}
Here, $N$ is the total number of sites of the square lattice, while
$q_i$ and $p_i$ represent generalized position and momentum operators
at a lattice site $i$, respectively, such that $[q_i,p_j]=i\,
\delta_{ij}$, with $i,j=1,\ldots, N$. The $N\times N$ matrices $A$ and
$B$ are real, symmetric, and semipositive; in the most general case,
they do not commute. The magnitude of quantum fluctuations are
characterized by the mean-square deviation of the average value of
these operators in the ground state: ${\bar q^2} = \sum_i \langle
q_i^2\rangle/N$ and ${\bar p^2} = \sum_i \langle p_i^2 \rangle/N$. As
mentioned in Sec. \ref{sec:magnetization}, if quantum fluctuations are
unbounded at the percolating regime, then these mean values diverge
and LRO is not possible, implying that order has to be destroyed
before $x_c$ is reached; that is, a quantum critical point should
exist before the percolation threshold. Here we show that this is not
the case and that LRO can persist up to and including $x_c$. As for
the spin Hamiltonians approximated by Eq. (\ref{eq:coupledHO}), the
exact disappearance of magnetic LRO may also depend on the spin value.

The eigenfrequencies of the harmonic oscillators of
Eq. (\ref{eq:coupledHO}) can be shown to be those of
\begin{equation}
\label{eq:omega}
\Omega_L^2=AB \quad \text{or} \quad \Omega_R^2=BA.
\end{equation}
Since, in general, $[A,B]$ is nonzero, the matrices $\Omega_L^2$ and
$\Omega_R^2$ are non-Hermitian and distinct. However, it is simple to
show that they have exactly the same real eigenvalues, $\omega^2_n$,
for $n=1,\dots, N$. This is also true for the matrices $\Omega_C^2 =
B^{1/2}AB^{1/2}$ and $\tilde\Omega_C^2 = A^{1/2}BA^{1/2}$.  Therefore,
the fluctuations of the position and momentum, averaged over all
sites, can be written as\cite{obs}
\begin{eqnarray}
{\bar q^2} & = & \frac{1}{2N}\; \text{tr} \left( B\; {\Omega_C}^{-1}
\right) \le \frac{b_\ast}{2}\, \kappa,
\label{eq:fluc-x}
\\ {\bar p^2} & = & \frac{1}{2N}\; \text{tr} \left( A\;
\tilde{\Omega_C}^{-1} \right) \le \frac{a_\ast}{2} \, \kappa,
\label{eq:fluc-p}
\end{eqnarray}
where $a_\ast$ and $b_\ast$ denote the maximum eigenvalues of the
semipositive definite matrices $A$ and $B$, respectively, and we used
that
\begin{eqnarray}
\label{eq:kappa}
\kappa = \frac{1}{N}\;\text{tr}\; {\Omega^{-1}_C}=
\frac{1}{N}\;\text{tr}\; {\tilde\Omega_C}^{-1} =
\int_0^{\omega_\text{max}} d\omega\; {\rho(\omega)}\, {\omega}^{-1},
\end{eqnarray}
since $\Omega_C$ and $\tilde\Omega_C$ share the same eigenvalues
($\omega_\text{max}$ being the largest), with spectral density
$\rho(\omega)$. Thus, the finiteness of the quantum fluctuations
reduces to the problem of the convergence of the integral in
Eq. (\ref{eq:kappa}) (quantum mechanics requires that ${\bar q^2}\,
{\bar p^2} \geq 1/4$). Notice that the quantum fluctuation correction
to the order parameter per unit of lattice site can be written as
\begin{equation}
\delta m_z = \alpha\, {\bar q^2}+ \beta\, {\bar p^2} + \gamma,
\end{equation}
where $\alpha$, $\beta$, and $\gamma$ are constants that depend on the
particular model and order parameter under consideration.

We now turn to applying these general results to specific bosonic
models based on Heisenberg Hamiltonians.

%%%%%%%%%%%%%%%%%%%%%%%%%%%%%%%%%%%%%%%%%%%%%%%%%%%%%%%%%%%%%%%%%%%%%%%%%%%%
\subsection{O(2) model}

This model is realized, for example, in an array of Josephson
junctions, where the variables $q_i$ correspond to the linearized
phase of the superconductor order parameter.\cite{senthil,Ali-Moore}
The matrices in Eq. (\ref{eq:coupledHO}) take the simple forms $A_{ij}
= K_{ij} -\Delta_{ij}$ and $B_{ij} = (U/J)\, \delta_{ij}$, where
$\Delta_{ij} = 1\, (0)$ when the nearest-neighboring sites $i,j$ are
both occupied (otherwise) and $K_{ij} = \delta_{ij}\, \sum_{l=1}^N
\Delta_{il}$, i.e., $K_{ii}$ counts the number of nearest neighbors of
site $i$. Here, $J$ denotes the Josephson coupling and $U$ is the
island charging energy (usually, $J\gg U$). The same structure occurs
in the case of vibrations in a diluted lattice.\cite{nakayama94} For
the O(2) model, the frequency eigenvalues are obtained from $\Omega^2
= (U/J)\, A$. The problem of determining the density of states of the
connectivity matrix $A$ can be mapped onto a problem of diffusion in a
disordered lattice.\cite{nakayama94} It can be shown that the density
of states of $\Omega$ goes $\rho(\omega)\propto \omega^{d^\ast-1}$. Up
to the percolation threshold, $d^\ast = d = 2$, while $d^\ast =
\tilde{d} \approx 4/3$ exactly at $x=x_c$. Therefore, the
site-averaged fluctuations are bounded, and the linear approximation
(as well as order) is maintained as long the ratio $J/U$ is large
enough.

%%%%%%%%%%%%%%%%%%%%%%%%%%%%%%%%%%%%%%%%%%%%%%%%%%%%%%%%%%%%%%%%%%%%%%%%%%%%%%
\subsection{Heisenberg antiferromagnetic model}

When we take $J_x=J_y=J_z$, the matrices take the form $A_{ij} =
K_{ij} - \Delta_{ij}$ and $B_{ij} = K_{ij} + \Delta_{ij}$. It is
useful to define a matrix $\Lambda = \Lambda^\top$ that has the effect
of changing the signs of $q_i$ and $p_i$ for all sites $i$ in one of
the two sublattices. Thus, $B = \Lambda A \Lambda$, and $\Omega_R^2 =
\Lambda A \Lambda A$. Notice that $\Gamma = \Lambda A$, is not
semipositive definite (in fact it is non-Hermitian), but has real
eigenvalues with the same magnitude as those of $\Omega$.

In the O(2) model, because $\Omega^2 \propto A$, one can directly
relate the energy eigenvalues of $\Omega$ to those of the matrix
$A$. In the AFM case, we need to obtain the density of states of
$\Gamma = \Lambda A$, which is not simply related to that of $A$
(since $[\Lambda,A] \ne 0$ in general for diluted systems). This
nontrivial relation between the eigenvalues of $\Omega$ and $A$ is
generally present in bosonic problems. For example, a similar problem
also appears in the work of Gurarie and Chalker in the relation
between their stiffness and frequency
eigenvalues.\cite{Gurarie-Chalker}

The problem of determining the density of states of $\Gamma$ for a
random dilution problem is one of the interesting open questions
related to the important difference between random fermionic and
bosonic systems. In a fermionic problem this question would be already
answered by matching the symmetries of $\Gamma$ to the Cartan
classification table.\cite{Zirnbauer} However, one cannot substitute
the matrix $A$ by an arbitrary random matrix with similar symmetries,
since that would violate the semi-positive definite constraint. In
this work we do not attempt to analytically resolve this problem;
however, we find numerical evidence that the density of states of
$\Gamma^2$ follows that of $A$ at low energies.

Our simulations show that the site-averaged fluctuations $\delta
m_i^z$ are bounded, both below and at the percolation threshold. This
means that order should exist up to and including the critical
dilution $x_c$, as long as $\delta m_i$ is small compared to the value
of the spin $S$. Recalling Eq. (\ref{eq:clustercontr}), we conclude
that there could be a minimum value $S_\text{min}$ for the spin, below
which there is a quantum phase transition for dilutions $x<x_c$. An
effective local spin smaller than $1/2$ can be realized in a bilayer
system with antiferromagnetic interlayer coupling.\cite{doublelayer}

%%%%%%%%%%%%%%%%%%%%%%%%%%%%%%%%%%%%%%%%%%%%%%%%%%%%%%%%%%%%%%%%%%%%%%%%%
\subsection{XXZ model}

In this case, $J_x=J_y \neq J_z$. We then have $A_{ij} = K_{ij} -
\Delta_{ij}$ and $B_{ij} = K_{ij} - \gamma \Delta_{ij}$, where
$\gamma$ measures the anisotropy. Alternatively, we can write $B =
(1+\gamma)/2 A + (1-\gamma)/2 \Lambda A \Lambda$, and so $\Omega^2 =
(1+\gamma)/2 A^2 + (1-\gamma)/2 \Lambda A \Lambda A$ (notice that for
$\gamma\to -1$ we have the same problem as the AFM). The analysis is
similar to the previous two cases. The amount of fluctuations is
controlled by the anisotropy, and it can be shown to be bounded if the
density of states follows that of the AFM case.

%%%%%%%%%%%%%%%%%%%%%%%%%%%%%%%%%%%%%%%%%%%%%%%%%%%%%%%%%%%%%%%%%%%%%%%%%
%%%%%%%%%%%%%%%%%%%%%%%%%%%%%%%%%%%%%%%%%%%%%%%%%%%%%%%%%%%%%%%%%%%%%%%%%
\section{Discussion and Conclusions}
\label{sec:conclusions}

In this work we have studied the role played by site dilution in
enhancing quantum fluctuations in the ground state of the Heisenberg
antiferromagnet in a square lattice. Using the linear spin-wave
approximation for this model, we have performed exact numerical
diagonalizations for lattices up to $36\times 36$, with dilution
fractions going from the clean limit to the classical percolation
threshold. Our results indicate a progressive, nonuniform shift of
spectral weight in the spin-wave excitation spectrum from high and to
low frequency as the dilution increases, with the high-frequency part
becoming more localized. The higher density of low-frequency, long
wavelength excitations leads to strong quantum fluctuations and a
decrease in the magnitude of the staggered magnetization. For
dilutions very close to the classical percolation threshold, we have
found that quantum fluctuations are sufficiently strong to nearly
match the clean-limit magnitude of the magnetization when $S=1/2$, but
not for higher spins. This is consistent with recent
neutron-scattering experiments with the $S=1/2$ dilute Heisenberg
antiferromagnet La$_2$Cu$_{1-x}$(Zn,Mg)$_x$O$_4$, which show that
long-range order disappears at around the classical percolation
threshold. However, quantum Monte Carlo simulations suggest that
quantum fluctuations should remain small and that the destruction of
long-range order is controlled only by the disappearance of the
infinite connected cluster. We understand this discrepancy between the
quantum Monte Carlo results and the linear spin-wave theory near the
classical percolation threshold as an indication that the latter has
its validity limited, as the magnitude of the order parameter is too
small.

While perhaps not quantitative-accurate, our simulations do allow us
to probe into the nature of the low-lying excited states of the dilute
antiferromagnet. We observe two clear humps in the density of states
at frequencies $\omega/JS= 2$ and 3. By calculating the
ensemble-averaged dynamical structure factor, we were able to
associate the appearance of these humps with the breaking of the
clean-limit magnon branch into three distinct but broad branches. The
new branches tend to be strongly localized (nondispersive) at high
frequencies and have a diffusive rather than ballistic nature at low
frequencies. In the literature, the multipeak structure had been
associated with the appearance of fractons in the excitation spectrum
as the dilution increases. From our simulations, it seems that the
picture is somewhat more complex. Besides the overall broadening, the
position of the high-frequency branch remains close to the
corresponding portion of original clean-limit magnon branch, while the
magnon velocity (the $d\omega/dq$ slope) in the low-frequency branch
is continuously reduced with increasing dilution. Therefore, it
appears that the fracton character also contaminates the low-frequency
branch. However, the lack of resolution due to the finite size of our
lattices does not allow for a conclusive picture. We did not attempt,
however, to study fracton states which can possibly occur above the
percolation threshold.\cite{alexander82,polatsek89}

It is important to remark that, in principle, due to Anderson
localization in two dimensions, we expect that in the infinite system
all excitations should in fact be localized for any finite
dilution. In order to probe more carefully strong localization and the
consequent exponential decay in real space, we need not only much
larger lattices, but also to calculate two-point correlators, which
goes beyond the applicability of linear spin-wave approximation. For
that same reason, we were not able to evaluate quantities such as the
spin stiffness, which involves matrix elements of higher than bilinear
operators.

We also studied the question of local quantum fluctuations as a way to
destroy long-range order. For finite-size lattices we found that the
weak links do not show strong quantum renormalizations. That provides
some indication that local quantum fluctuations may not be sufficient
to change the dominance of the classical percolation picture.

Using a more general analytical formulation, we have argued that there
exists an upper bound for the quantum fluctuations in any model with a
classically ordered ground state whose Hamiltonians can be mapped onto
that of a system of coupled harmonic oscillators. The amount of
quantum fluctuations depends directly on the low-energy behavior of
the density of states of the associated bosonic problems. Our exact
diagonalization of the linear spin-wave Hamiltonian on a percolating
lattice led us to identify the value of the upper bound for one
particular type of model and can readily be used to find similar
values for any other bosonic model. This could be used to study a
large class of spin Hamiltonians that can be bosonized in the ordered
phase.

%%%%%%%%%%%%%%%%%%%%%%%%%%%%%%%%%%%%%%%%%%%%%%%%%%%%%%%%%%%%%%%%%%%%%%%%%
%%%%%%%%%%%%%%%%%%%%%%%%%%%%%%%%%%%%%%%%%%%%%%%%%%%%%%%%%%%%%%%%%%%%%%%%%
\begin{acknowledgments}

We thank I. Affleck, H. Baranger, J. Chalker, A. Chernyshev,
M. Greven, J. Moore, C. Mudry, A. Sandvik, T. Senthil, O. Sushkov,
O. Vajk, A. Vishwanath, and M. Vojta for useful conversations. We
acknowledge financial support from CNPq and PRONEX in Brazil (E.R.M.),
and from the NSF through grants No. DMR-0103003 (E.R.M.),
No. DMR-0343790 (A.H.C.N.), and No. DMR-0305482 (C.C.).

\end{acknowledgments}

%%%%%%%%%%%%%%%%%%%%%%%%%%%%%%%%%%%%%%%%%%%%%%%%%%%%%%%%%%%%%%%%%%%%%%%%%
%%%%%%%%%%%%%%%%%%%%%%%%%%%%%%%%%%%%%%%%%%%%%%%%%%%%%%%%%%%%%%%%%%%%%%%%%

\appendix

%%%%%%%%%%%%%%%%%%%%%%%%%%%%%%%%%%%%%%%%%%%%%%%%%%%%%%%%%%%%%%%%%%%%%%%%%
\section{Formulation in terms of coupled harmonic oscillator}
\label{sec:appendixA}

The spin-wave Hamiltonian of Eq. (\ref{eq:HSW}) can be represented in
terms of position and momentum operators. In this language, it becomes
more transparent that the problem of finding the eigenvalues and
eigenvectors of the Hamiltonian can be solved by the diagonalization
of two real symmetric matrices, an alternative to the non-Hermitian
eigenvalue formulation of Sec. \ref{sec:nonhermitian}.

Let us perform the following operator transformation:
\begin{equation}
q_i = \frac{a_i + a_i^\dagger}{\sqrt{2}} \qquad {\rm and} \qquad p_i =
\frac{a_i - a_i^\dagger}{i\sqrt{2}},
\end{equation}
for all $i=1,\ldots,N$. Notice that the operators $q_i$ and $p_i$ obey
the canonical position-momentum commutation relations. It is
convenient to adopt a matrix formulation for the problem, namely,
\begin{equation}
H = \frac{JS}{2} \left( q^T {\bf A}\, q + p^T {\bf B}\, p \right),
\end{equation}
where $x = \{x_i\}_{i=1\ldots N}$ and $p = \{p_i\}_{i=1\ldots N}$
denote vectors of position and momentum operators, while $[{\bf
A}]_{ij} = K_{ij} + \Delta_{ij}$, and $[{\bf B}]_{ij} = K_{ij} -
\Delta_{ij}$. The quantum fluctuation correction to the sublattice
magnetization can be written as
\begin{equation}
\delta m = \frac{1}{2N} \text{tr} \left[ \langle q\, q^T \rangle +
\langle p\, p^T \rangle \right] - \frac{1}{2}.
\end{equation}
We can diagonalize ${\bf B}$ through an orthogonal transformation
${\bf U}$,
\begin{equation}
{\bf U}^T {\bf B}\, {\bf U} = {\bf b} \qquad \text{(diagonal)}
\end{equation}
and define new coordinates such that
\begin{equation}
q = {\bf U}\, q^\prime \ \ \text{and} \ \ 
p = {\bf U}\, p^\prime.
\end{equation}
Defining ${\bf A}^\prime = {\bf U}^T {\bf A}\, U$, we then have that
\begin{equation}
H = \frac{JS}{2} \left( q^{\prime\, T} {\bf A}^\prime\, q^\prime +
p^{\prime\, T} {\bf b}\, p^\prime \right)
\end{equation}
and
\begin{equation}
\delta m = \frac{1}{2N} \left[ \text{tr} \left( \left\langle
q^\prime\, q^{\prime\, T} \right\rangle \right) + \text{tr} \left(
\left\langle p^\prime\, p^{\prime\, T} \right\rangle \right) \right] -
\frac{1}{2}.
\end{equation}

It is not difficult to prove that all elements in the diagonal of
${\bf b}$ are positive except one, $b_1$, which is zero. In order to
eliminate this zero mode, we subtract the corresponding row or line in
all vectors and matrices, which amounts to a reduction in the Hilbert
space (or, alternatively, to set $q^\prime_0 = 0$): $\{q^\prime\}_N
\rightarrow \{q^\prime\}_{N-1}$ and $\{p^\prime\}_N \rightarrow
\{p^\prime\}_{N-1}$. Also, $[{\bf A}^\prime]_{N\times N} \rightarrow
[{\bf A}^\prime]_{N-1\times N-1}$ and $[{\bf b}]_{N\times N}
\rightarrow [{\bf b}]_{N-1\times N-1}$. Notice that now all $b_k>0$,
$k=2,\ldots N$. Thus, in this new space, we can perform the following
rescaling:
\begin{equation}
q^{\prime\prime} = {\bf b}^{-1/2}\, q^\prime \ \ \text{and} \ \ 
p^{\prime\prime} = {\bf b}^{1/2}\, p^\prime,
\end{equation}
which allows us to write
\begin{equation}
H = \frac{JS}{2} \left( q^{\prime\prime\, T} {\bf A}^{\prime\prime}\,
q^{\prime\prime} + p^{\prime\prime\, T}\, p^{\prime\prime} \right),
\end{equation}
with
%
%\begin{equation}
${\bf A}^{\prime\prime} = {\bf b}^{1/2} {\bf A}^\prime\, {\bf b}^{1/2}$
%\end{equation}
%
as well as
\begin{eqnarray}
\delta m & = & \frac{1}{2N} \left[ \text{tr} \left( {\bf b}\, \left
\langle q^{\prime\prime}\, q^{\prime\prime\, T} \right\rangle \right)
+ \text{tr} \left( {\bf b}^{-1}\, \left\langle p^{\prime\prime}\,
p^{\prime\prime\, T} \right\rangle \right) \right. \nonumber \\ & &
\left. + \langle p_1^{\prime\, 2} \rangle \right] - \frac{1}{2},
\end{eqnarray}
where the last term within the square brackets reflects the existence
of a zero mode. It is useful now to return to the site basis by
carrying out the inverse rotation,
\begin{equation}
q^{\prime\prime\prime} = {\bf U}\, q^{\prime\prime} \ \ \text{and} \ \ 
p^{\prime\prime\prime} = {\bf U}\, p^{\prime\prime},
\end{equation}
such that
\begin{equation}
H = \frac{JS}{2} \left( q^{\prime\prime\prime\, T} {\bf C}\,
q^{\prime\prime\prime} + p^{\prime\prime\prime\, T}\,
p^{\prime\prime\prime\,} \right),
\end{equation}
where
%
%\begin{equation}
${\bf C} = {\bf U}\, {\bf A}^{\prime\prime}\, {\bf U}^T$
%\end{equation}
%
is the new connectivity matrix. If we define
\begin{equation}
{\bf B}^{1/2} = {\bf U}\, {\bf b}^{1/2}\, {\bf U}^T \ \ \text{and} \ \
{\bf B}^{-1/2} = {\bf U}\, {\bf b}^{-1/2}\, {\bf U}^T
\end{equation}
within the subspace $N-1 \times N-1$ where no zero mode is present, we
can write the connectivity matrix as
\begin{equation}
{\bf C} = {\bf B}^{1/2}\, {\bf A}\, {\bf B}^{1/2}.
\end{equation}
Also, notice that the inverse rotation does not change the expression
of the sublattice magnetization,
\begin{eqnarray}
\delta m & = & \frac{1}{2N} \left[ \text{tr} \left( {\bf b}^{-1}\,
\left\langle q^{\prime\prime\prime}\, q^{\prime\prime\prime\, T}
\right\rangle \right) + \text{tr} \left( {\bf b}\, \left\langle
p^{\prime\prime\prime}\, p^{\prime\prime\prime\, T} \right\rangle
\right) \right. \nonumber \\ & & \left. + \langle p_0^{\prime\, 2}
\rangle \right] - \frac{1}{2}.
\end{eqnarray}

We can now perform the last operation, namely, the diagonalization of
the connectivity matrix,
\begin{equation}
q^{iv} = {\bf V}\, q^{\prime\prime\prime} \ \ \text{and} \ \ 
p^{iv} = {\bf V}\, p^{\prime\prime\prime},
\end{equation}
yielding
\begin{equation}
H = \frac{JS}{2} \left( q^{iv\, T} {\bf c}\, q^{iv} + p^{iv\, T}\,
p^{iv\,} \right),
\end{equation}
where
\begin{equation}
{\bf V}^T {\bf C}\, {\bf V} = {\bf c} \ \ \  \text{(diagonal)},
\end{equation}
with all $c_n>0$, $n=1,\ldots N-1$. We finally arrive to a system of
decoupled harmonic oscillators. The quantum fluctuation part of the
magnetization becomes
\begin{eqnarray}
\delta m & = & \frac{1}{2N} \left[ \text{tr} \left( {\bf b}^{-1}\,
{\bf V}^T\, \left\langle q^{iv}\, q^{iv\, T} \right\rangle \, {\bf V}
\right) \right. \nonumber \\ & & \left. + \text{tr} \left( {\bf b}\,
{\bf V}^T\, \left\langle p^{iv}\, p^{iv\, T}\, \right\rangle {\bf V}
\right) + \langle p_0^{\prime\, 2} \rangle \right] - \frac{1}{2}.
\end{eqnarray}
However, we know that
\begin{equation}
[\langle q^{iv}\, q^{iv\, T} \rangle]_{nm} = \delta_{nm}\, \langle
q_n^{iv\, 2} \rangle = \frac{1}{2} [ {\bf c} ]_{nm}^{-1/2}
\end{equation}
and
\begin{equation}
[\langle p^{iv}\, p^{iv\, T} \rangle]_{nm} = \delta_{nm}\, \langle
p_n^{iv\, 2} \rangle = \frac{1}{2} [ {\bf c} ]_{nm}^{1/2}.
\end{equation}
Therefore,
\begin{eqnarray}
\delta m & = & \frac{1}{4N} \left[ \text{tr} \left( {\bf b}^{-1}\,
{\bf V}^T\, {\bf c}^{-1/2}\, {\bf V} \right) + \text{tr} \left( {\bf
b}\, {\bf V}^T\, {\bf c}^{1/2}\, {\bf V} \right) \right. \nonumber \\
& & \left. + \langle p_0^{\prime\, 2} \rangle \right] - \frac{1}{2}.
\end{eqnarray}

It is interesting to notice that, since ${\bf A} = {\bf O}\, {\bf B}\,
{\bf O}$, we have that
\begin{equation}
{\bf C} = {\bf \Omega}^2,
\end{equation}
where
\begin{equation}
{\bf \Omega} = {\bf B}^{1/2}\, {\bf O}\, {\bf B}^{1/2}.
\end{equation}
Thus, it is easy to see that if ${\bf C}$ were a positive matrix, then
we would be able to write ${\bf V}^T\, {\bf c}^{1/2}\, {\bf V} \equiv
{\bf C}^{1/2} = {\bf \Omega}$. That would allow us to simplify the
expression for the sublattice magnetization a step further. However,
the connectivity matrix is not necessarily positive.

%%%%%%%%%%%%%%%%%%%%%%%%%%%%%%%%%%%%%%%%%%%%%%%%%%%%%%%%%%%%%%%%%%%%%%%%%
\section{Numerical diagonalization of the non-Hermitian matrix}
\label{sec:appendixB}

The diagonalization of the non-Hermitian matrices consists of five
steps. First, the real (but asymmetric) matrix $M^{(+)}$ is reduced to
an upper Hessenberg form through an orthogonal transformation, namely,
$A = Q M^{(+)} Q^T$. This is done by using the LAPACK subroutines
DGEHRD and DORGHR. Second, we use the LAPACK subroutine DHSEQR to
perform the Schur factorization of the Hessenberg matrix: $A = Z T
Z^T$. That allows us to obtain the eigenvalues and the Schur vectors,
which are contained in the orthogonal matrix $Z$. Third, using another
LAPACK subroutine, DTREVC, we extract from $Z$ both right and left
eigenvectors of $M^{(+)}$. In the fourth step we renormalize all
eigenvectors $\{\phi^{(\pm)}\}$ such that they satisfy
Eq. (\ref{eq:newsumrule1}) [the condition in
Eq. (\ref{eq:newsumrule2}) is automatically satisfied], sort the
eigenvalues $\{\lambda_n\}$ in ascending order, and extract the zero
mode from the spectrum. For some realizations, the lowest eigenvalues
next to the zero mode cannot be distinguished from the zero mode
itself and are therefore neglected. Finally, the correct linear
combination of $\phi_{n}^{(\pm)}$ that provides the correct $u_{n}$
and $v_{n}$ for each positive eigenfrequency $\omega_n =
\sqrt{\lambda_n}$ is obtained according to the algorithm presented in
Sec. \ref{sec:simplify}.

%%%%%%%%%%%%%%%%%%%%%%%%%%%%%%%%%%%%%%%%%%%%%%%%%%%%%%%%%%%%%%%%%%%%%%%%%
%%%%%%%%%%%%%%%%%%%%%%%%%%%%%%%%%%%%%%%%%%%%%%%%%%%%%%%%%%%%%%%%%%%%%%%%%

%%%%%%%%%%%%%%%%%%%%%%%%%%%%%%%%%%%%%%%%%%%%%%%%%%%%%%%%%%%%%%%%%%%%%%%%%


\begin{thebibliography}{11}
\bibitem{obs} When defining the inverse matrices in
Eqs. (\ref{eq:fluc-x}) and (\ref{eq:fluc-p}), the subspaces
corresponding to zero eigenvalue states have been eliminated.
\end{thebibliography}
\end{document}